\documentclass[showpacs,aps,pra,floatfix,reprint,superscriptaddress,footinbib,citeautoscript]{revtex4-2}
\usepackage{xfrac}
\usepackage{amssymb}
\usepackage{braket}
\usepackage{graphicx}
\usepackage{verbatim}
\usepackage{etoolbox}
\usepackage{amsfonts}
\usepackage{amsmath}
\usepackage[utf8]{inputenc}
\usepackage{mathtools}
\usepackage{soul,xcolor,colortbl}
\usepackage{hyperref} \hypersetup{colorlinks=true,citecolor=blue,linkcolor=blue,urlcolor=black}
\usepackage{morefloats}
\usepackage{natmove}

\usepackage{bm}
\usepackage{array}
\newcolumntype{C}[1]{>{\centering\arraybackslash}p{#1}}\usepackage{soul}
\usepackage{color}
\def\AFLOW{{\small AFLOW}}
\def\AFLOWorg{{\sf \AFLOW.org}}

\def\GFA{{\small GFA}}

\def\AFLOWPOCC{{\small AFLOW-POCC}}

\def\AFLOWAPL{{\small AFLOW-APL}}
\def\RESTAPI{{\small REST-API}}
\def\AFLUX{{\small AFLUX}}

\def\GQCA{{\small GQCA}}
\def\ATAT{{\small AT-AT}}

\def\DHNN{{\Delta H \left[N|\left\{1,\cdots\!,N\!\!-\!\!1\right\}\right]}}

\begin{document}
\title{\Large AFLOW for alloys}
\author{Cormac Toher}
\email[]{cormac.toher@utdallas.edu}
\affiliation{Department of Materials Science and Engineering, The University of Texas at Dallas, Richardson, Texas 75080, USA}
\affiliation{Department of Chemistry and Biochemistry, The University of Texas at Dallas, Richardson, Texas 75080, USA}
\author{Stefano Curtarolo}
\email[]{stefano@duke.edu}
\affiliation{Materials Science, Electrical Engineering, and Physics, Duke University, Durham NC, 27708}

\date{\today}

\begin{abstract}
\noindent
Many different types of phases can form within alloys, from highly-ordered intermetallic compounds, to structurally-ordered but chemically-disordered solid solutions, and structurally-disordered (i.e. amorphous) metallic glasses.
The different types of phases display very different properties, so predicting phase formation is important for understanding how materials will behave.
Here, we review how first-principles data from the \AFLOW\ repository and the {\sf aflow++} software can be used to predict phase formation in alloys, and describe some general trends that can be deduced from the data, particularly with respect to the importance of disorder and entropy in multicomponent systems.
\end{abstract}

\maketitle

\section*{Introduction}
\begin{figure}[ht]
  \includegraphics[width=0.8\columnwidth]{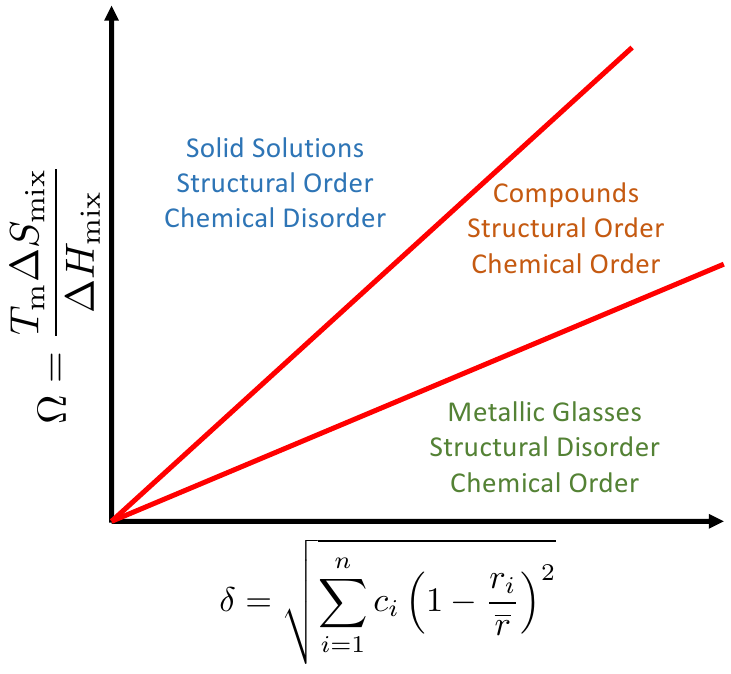}
  \vspace{-2mm}
  \caption{\small {\bf Order and disorder in alloy space.}
    Phases formed in alloys range from structurally disordered metallic glasses (formed when components have a large range of ionic radii --- descriptor $\delta$),  to ordered intermetallic compounds, and solid solutions (formed when components have a large ratio of mixing entropy $\Delta S_{\mathrm{mix}}$ to mixing enthalpy $\Delta H_{\mathrm{mix}}$).
  Figure is inspired by Figure 1 in Ref. \onlinecite{Yang_MCP_Entropy_2012}.}
  \label{fig:disorder_space}
\end{figure}

\noindent
A wide variety of phases can form within alloys \cite{massalski}, from highly-ordered intermetallic compounds \cite{curtarolo:art20}, to structurally-ordered and chemically-disordered solid solutions \cite{Gau2015high, George_NRM_2019, curtarolo:art183, Miracle_HEA_NComm_2019, Miracle_HEAs_NComm_2015}, and structurally-disordered (i.e. amorphous) metallic glasses \cite{greer1993confusion, Greer_metallic_glasses_review_2009, Johnson_BMG_2009, greer2015new, miracle2004structural} (Figure \ref{fig:disorder_space}).
Different phase-types lead to very different properties: intermetallics are hard but brittle; solid solutions display solid-solution strengthening and can even overcome the strength-ductility trade-off \cite{HEAprop2, Li_SRep_HEA_Dual_Phase_2017}; while metallic glasses can be processed using thermoplastic blow-molding \cite{Schroers_blow_molding_2011}.
Metallic engineering materials typically contain multiple phases forming a carefully-engineered microstructure: small plates or needles of harder intermetallic or ceramic compounds (e.g. cementite (Fe$_3$C) in steels, or the gamma phase in Ni-superalloys) embedded in a softer, more ductile matrix.
The dispersal of the intermetallic phases in the matrix is controlled by factors such as the annealing time and quench rate, which limits the extent of diffusion, and can lead to materials with good combinations of strength and toughness.
Other microstructures can be formed by rapid unmixing processes such as spinodal decomposition \cite{Cahn1961795, Cahn1962179}, where the internal stresses at the interface between phases impede the motion of dislocations, increasing hardness \cite{Cahn_ActaMetal_1963}.

Due to the large effect of phase composition on alloy properties, predicting which phases are most likely to form under specific conditions, and which processing pathways can give rise to desired combinations of phases, is an important challenge. Experimental investigation of phase formation can be expensive and slow, due to the time needed for atoms to diffuse into the equilibrium microstructure.
Therefore, computational methods have been developed to predict phase formation, including CALPHAD (calculation of phase diagrams) \cite{Kaufman_CALPHAD_1970, CoFeMnNi,calph1,Miracle_HEAs_NComm_2015} that interpolates/extrapolates from existing experimental data \cite{Thermocalc_software, COST_alloy_database}, molecular dynamics simulations using either empirical force-fields or ab initio calculations \cite{kresse_vasp}, and first-principles calculations of formation enthalpies that can be combined with cluster expansion \cite{deFontaine_ssp_1994, atat3, atat4} or machine-learning \cite{curtarolo:art174} to generate large structure-energy ensembles for use in Monte Carlo simulations \cite{widom2015high,widom1,widom2} or statistical thermodynamics models \cite{curtarolo:art139}.

\begin{figure*}[ht]
  \includegraphics[width=0.98\textwidth]{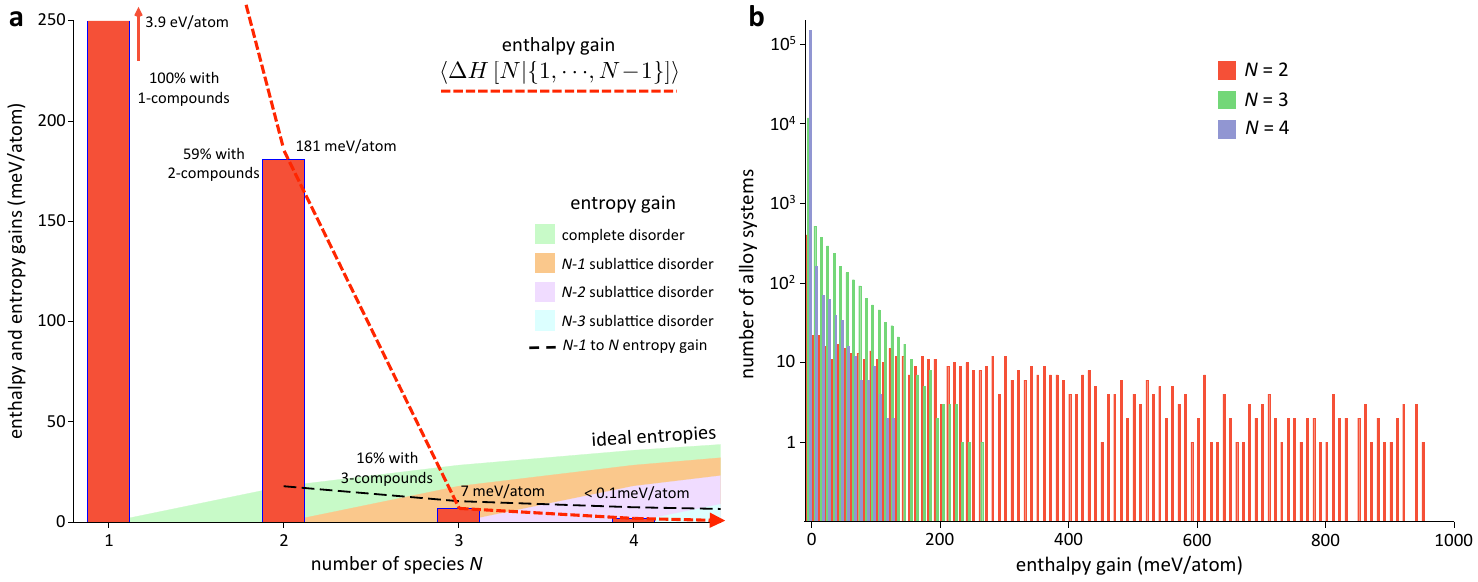}
  \vspace{-2mm}
  \caption{\small
    {\bf Enthalpy and entropy gains.}
    ({\bf a}) Expectation value of enthalpy gain $\left<\DHNN\right>$ and ideal entropy contributions as functions of the number of elements, $N$.
    Even at room temperature, entropy eventually overwhelms enthalpy in determining the phase stability of multi-component systems.
    ({\bf b}) Enthalpy gain distribution for $N=2,3,4$; gains decrease sharply with increasing number of elements.
  Figure is adapted from Figure 1 in Ref. \onlinecite{curtarolo:art152}.}
  \label{fig:inevitable_disorder}
\end{figure*}

The large data sets required to train cluster expansion or machine-learning models require automated management of first-principles calculations, such as that performed by \AFLOW\ \cite{curtarolo:art191, curtarolo:art65,curtarolo:art115,curtarolo:art125,curtarolo:art110,curtarolo:art87,curtarolo:art85,curtarolo:art63,curtarolo:art58,curtarolo:art57,curtarolo:art53,curtarolo:art49,curtarolo:art142}.
\AFLOW\ generates input files for density functional theory (DFT) calculations \cite{vasp_prb1996, kresse_vasp_paw, martin2004electronic} by decorating structural prototypes from the in-built Library of Crystallographic Prototypes \cite{curtarolo:art121, curtarolo:art145, curtarolo:art173}, monitors the calculations and automatically detects and corrects any errors, and finally extracts relevant information from the output files.
Finally, the extracted data is formatted and added to the \AFLOW\ database \cite{curtarolo:art190,curtarolo:art75,curtarolo:art92,curtarolo:art104}, which is accessible using the \AFLUX\ search-API and \AFLOW\ \RESTAPI\ \cite{curtarolo:art92, curtarolo:art128}.
The \AFLOW\ database currently contains over 3.5 million entries, covering binary (1738), ternary (30,289) and quaternary (150,659) systems.
Here, we review how \AFLOW\ data can be used to predict the formation of disordered materials, and to make general statements about the types of phases that will form in multi-component materials.

\section*{The inevitability of disorder}
\noindent
The central role of entropy in the formation of multi-component systems can be demonstrated using large computational data sets.
Analysis \cite{curtarolo:art152} of formation enthalpies for 202,261 binary, 974,808 ternary and 432,840 quaternary entries from the \AFLOWorg\ {\it ab-initio} materials repository with the associated tools \cite{curtarolo:art128} shows that the gain in formation enthalpy from adding additional elements decreases rapidly with increasing number of elements, while the configurational entropy continues to grow monotonically.

The formation enthalpy ``gain'' \cite{curtarolo:art152} $\DHNN$ of an $N$-element ordered compound with respect to combinations of $\{1,\cdots\!,N\!-\!1\}$-element ordered sub-components can be defined as the energetic distance of the enthalpy of the $N$-element compound, $H\left[N\right]$, below the $\{1,\cdots\!,N\!-\!1\}$ convex-hull hyper-surface
$H_{\rm hull} \left[\{1,\cdots\!,N\!\!-\!\!1\}\right]$ generated from its $\{1,\cdots\!,N\!\!-\!\!1\}$-element components:

\begin{widetext}
\begin{equation*}
 \DHNN \equiv \left\{ \begin{array}{ll}
              H_{\rm hull} \left[\{1,\cdots\!,N\!\!-\!\!1\}\right]-H\left[N\right] & \mbox{if $H\left[N\right]<H_{\rm hull} \left[\{1,\cdots\!,N\!\!-\!\!1\}\right]$};\\
             0 & \mbox{otherwise}.\end{array} \right.
\end{equation*}
\end{widetext}

For binary compounds, $\Delta H\left[2|1\right]$ is equivalent to the usual formation enthalpy; for $N>2$, the formation enthalpies can be written as sums of the recursive gains.
Physically, $\DHNN$ represents the enthalpy gained by creating an ordered $N$-species compound out of all the combinations of ordered $\left\{ 1-, \cdots\!, N\!-\!1\right\}$-ones.
The expected enthalpy gains can be extracted from the metal alloy phase diagrams generated with \AFLOWorg\ data.

The expectation value of the enthalpy gain $\left<\Delta H\right>$ is shown in Figure \ref{fig:inevitable_disorder}(a) for $N=2,3,4$, with the cohesive energy of the elemental reference phase shown for $N=1$.
$\left<\Delta H\right>$ is $\sim$181~meV/atom for binaries, reducing to $\sim$7~meV/atom for ternaries, and effectively disappearing ($< 0.1$~meV/atom) for quaternaries.
The frequency distributions of enthalpy gains for all systems are displayed in Figure \ref{fig:inevitable_disorder}(b).
588 (59\%) of binary systems show enthalpy gain, forming a total of 1995 binary compounds; 2237 (16\%) of ternaries form a total of
3040 ternary compounds, and just 426 (0.3\%) of quaternaries form quaternary compounds.
434 binary, 192 ternary but only 10 quaternary systems have an enthalpy gain exceeding 100~meV/atom.
Therefore, as the number of elements in an alloy increases, the likelihood of finding new materials consisting of single-phase intermetallic compounds drops rapidly.
Instead, multi-component materials are dominated by disorder, consisting of solid-solutions, metallic glasses and multiple separate phases \cite{Miracle_ActaMat_2017}.

\begin{figure*}
  \includegraphics[width=1.00\textwidth]{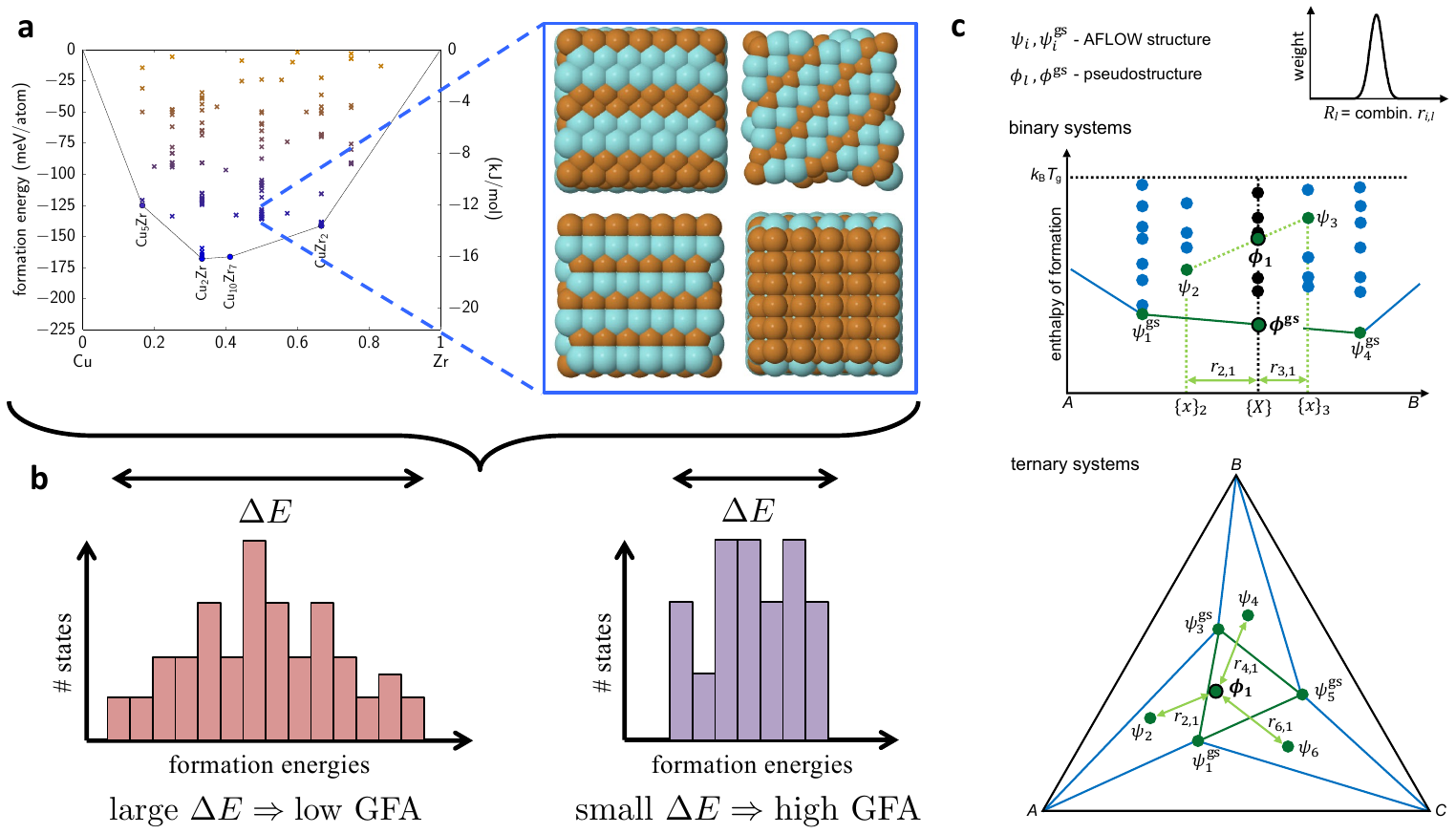}
  \vspace{-4mm}
  \caption{\small \textbf{Descriptor for glass forming ability.}
    The glass forming ability of metal alloy systems can be predicted from the
    spread of formation energies of relevant ordered structures.
    ({\bf a}) Different structural phases with similar energies compete against each other during solidification,
    frustrating crystallization and promoting glass formation.
    ({\bf b}) A broad distribution of energies implies a low glass forming ability,
    while a narrow distribution indicates a high glass forming ability.
    ({\bf c}) The GFA of intermediate stoichiometries can be estimated by generating pseudostructures $\phi$ that are linear combinations of the real structures $\psi$
    --- particularly useful for ternary alloys.
  Panels (a) and (b) adapted from Figure 1 of Ref. \onlinecite{curtarolo:art142}, and panel (c) adapted from Figure 1 of Ref. \onlinecite{curtarolo:art154}.
}
  \label{fig:GFA_descriptor}
\end{figure*}

While this analysis was  performed for alloys composed of 45 different metallic elements, this finding is expected to generalize to other systems such as high-entropy ceramics \cite{curtarolo:art99, curtarolo:art140, curtarolo:art183}, although formation enthalpy corrections \cite{curtarolo:art150, curtarolo:art172} to account for DFT inaccuracies will be required.
Mixing different types of bonding can generate additional enthalpic stabilization, increasing the threshold number of elements at which entropy dominates.
Meanwhile, reciprocal systems might be able to reduce the threshold, due to the additional entropic stabilization of every equivalent sublattice.

\section*{Glass formation}
\noindent
Metallic glasses are amorphous materials, lacking an ordered crystal lattice and its associated defects, providing them with a combination of excellent mechanical properties~\cite{chen2015does} and
plastic-like processability~\cite{schroers2006amorphous,Schroers_blow_molding_2011,kaltenboeck2016shaping},
rendering them of great interest for a range of commercial and industrial
applications~\cite{Schroers_Processing_BMG_2010,johnson2016quantifying,ashby2006metallic}.

\begin{figure*}
  \includegraphics[width=0.99\textwidth]{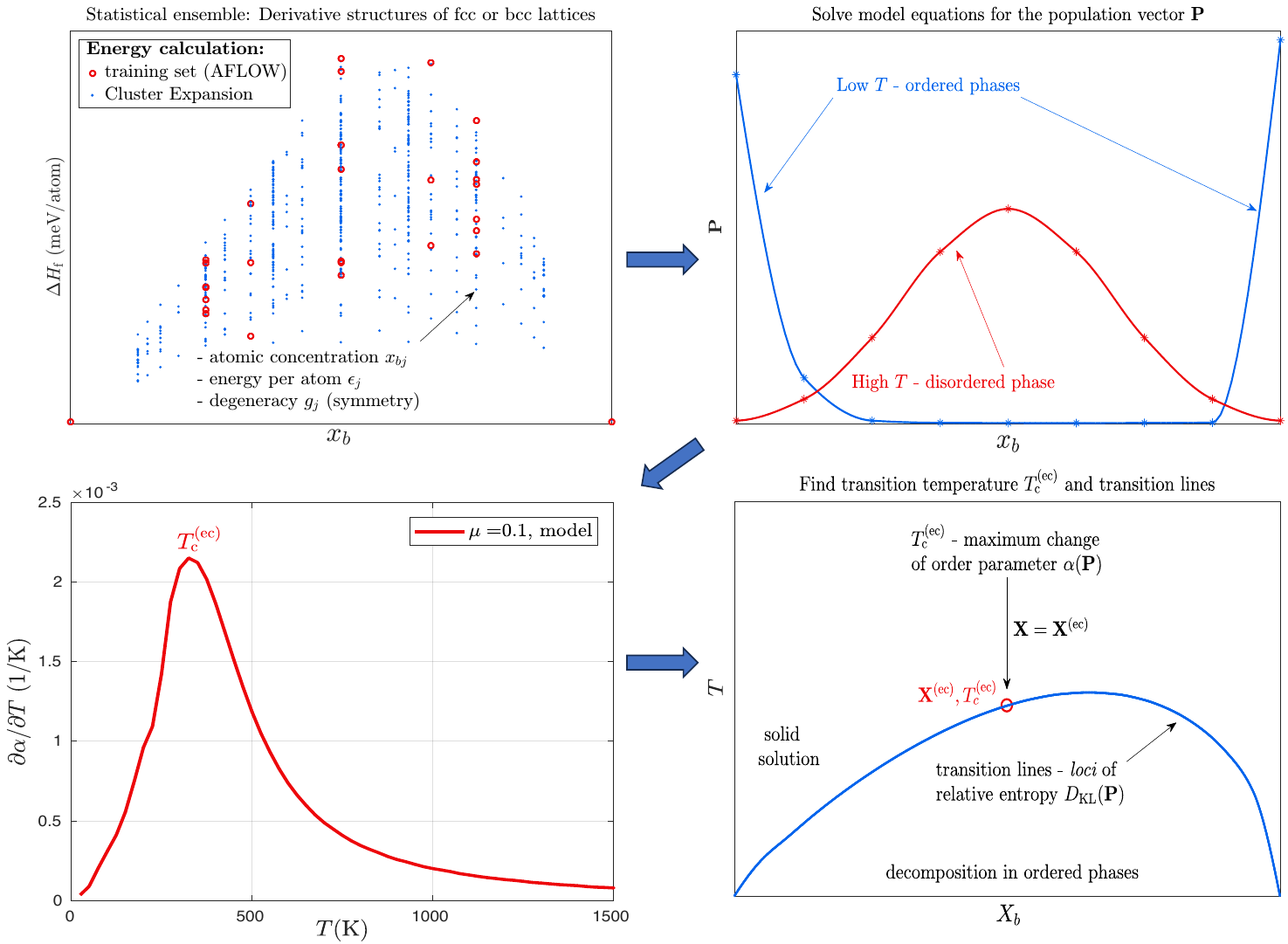}
  \vspace{-3mm}
  \caption{\small
    {\bf Outline of the method to predict order-disorder transitions.}
    ({\bf a})  Statistical ensemble is constructed: atomic configurations are generated and their multiplicity is calculated. Their energies are then estimated using cluster expansion with \AFLOW\ energies serving as the training set.
    ({\bf b}) The \GQCA\ model is solved for the population vector ${\bf P}({\bf X},T)$.
    ({\bf c}) $T_{\mathrm c}$ at equi-composition is evaluated as the maximum temperature derivative point of the order parameter $ \alpha({\bf X}^{\mathrm{(ec)}},T)$.
    ({\bf d}) Boundary lines are found assuming $D_{\mathrm{KL}}({\bf X},T_{\mathrm{c}}) \approx D_{\mathrm{KL}}({\bf X}^{\mathrm{(ec)}},T^{\mathrm{(ec)}}_{\mathrm{c}})$.
    Figure is adapted from Figures 1 and 2 in Ref. \onlinecite{curtarolo:art139}.
  }
\label{fig:ltvc}
\end{figure*}

A model based on the competition between different ordered structures has been successfully used to predict the glass forming ability (GFA) of metallic glasses \cite{curtarolo:art112, curtarolo:art154}.
The model is based on the ansatz that, at a particular composition, if an alloy system has a large number of structures that are similar to each other in energy, but different from each other structurally, then competition between the ordered phases during solidification will frustrate crystallization and promote glass formation (Figure  \ref{fig:GFA_descriptor}(a, b)).
Atomic environment~\cite{daams_villars:environments_2000,daams:cubic_environments} comparisons determine the similarity of ordered crystalline phases,
enabling the formulation of a quantitative descriptor that can be applied to the entire \AFLOWorg\ database.
The different structures are weighted according to a Boltzmann distribution to create the \GFA\ descriptor.
The model has been applied to \AFLOW\ data for binary and ternary alloy systems, and successfully predicts 73\% of the glass forming compositions for a set of 16 experimentally characterized binary alloy systems.
Notably, the model also predicts that 17\% of binary alloy systems should have a GFA greater than CuZr, a well-known glass-former.
Therefore, there are likely to be many more glass-forming alloy compositions discovered in the future.

A limitation of this model is the granularity of available calculated formation enthalpy data: calculations are performed for specific compositions that have structures with reasonably-sized primitive cells.
For compositions with 3 or more elements, this becomes a major issue, which can be addressed by also considering competition between structures with different stoichiometries.
Compositions with stoichiometries in-between those of common ordered structures can be considered as consisting of combinations of local structures with different compositions \cite{curtarolo:art154},
representing local stoichiometric fluctuations within the sample.
Pseudo-structures can then be constructed as linear combinations of the formation enthalpies and atomic environments of the real structures (Figure \ref{fig:GFA_descriptor}(c)).
The GFA descriptor can then be calculated using these pseudo-structures for intermediate stoichiometries, allowing prediction of glass formation across the entire composition space \cite{curtarolo:art154}.

\section*{Solid solutions and high-entropy alloys}
\noindent
In solid solutions, different elements occupy the same crystallographic site on a lattice. High-entropy alloys are solid solutions with multiple (usually at least 4 or 5) components forming a single phase \cite{Gau2015high, George_NRM_2019, HEAapp1,HEAapp2,HEAprop1,HEAprop2}, typically on a simple fcc or bcc lattice.
Similar to the method described above for predicting glass formation, materials that form high-entropy single-phases can be rapidly identified using a descriptor (EFA, \underline{e}ntropy-\underline{f}orming-\underline{a}bility) based on the distribution of energies for a set of ordered configurations based on a particular lattice \cite{curtarolo:art140}.
A narrow distribution of energies indicates miscibility leading to the formation of a high-entropy single phase, while a broader distribution corresponds to preferred ordered configurations leading to phase separation.

The EFA descriptor is qualitative, only indicating which compositions are most likely to form solid solutions.
Directly predicting the order-disorder transition temperatures from first-principles for alloy systems requires Monte Carlo simulations combined with \textit{ab initio} calculations \cite{widom2015high,widom1,widom2}, which can be computationally expensive.
To accelerate the prediction of phase diagrams, \AFLOW\ data was used to formulate a descriptor for the order-disorder transition temperature \cite{curtarolo:art139}, incorporating formation enthalpy calculations into a mean field statistical mechanics model, and making use of order parameters for predicting the transition temperature of a multi-component system into a solid solution phase.

The main steps for method to predict the transition temperature are \cite{curtarolo:art139}:

{\bf i.} \AFLOW\ data for binary alloy systems are leveraged to train
cluster expansion ({\small CE}) \cite{deFontaine_ssp_1994, sanchez_ducastelle_gratias:psca_1984_ce} models (Figure \ref{fig:ltvc}(a)), within
the \underline{A}lloy \underline{T}heoretic \underline{A}utomated \underline{T}oolkit (\ATAT) \cite{atat4} implementation.
The zero temperature energies $\varepsilon_j$ of atomic configurations based on derivative structures of  fcc and bcc lattices are then estimated using the {\small CE} models.

{\bf ii.} The atomic configurations are then incorporated into the  \underline{g}eneralized
\underline{q}uasi-\underline{c}hemical \underline{a}pproximation (\GQCA) mean field statistical mechanical model \cite{GQCA1, GQCA2}.
 \GQCA\ is particularly suitable for solid solutions, where long-range order is not expected to be important and the material is spatially homogeneous.

The thermodynamic potential $\Phi$ within the \GQCA\ model is given by
\begin{equation} \label{FE}
  \Phi = N \cdot \left( \sum_{j} \varepsilon_j P_j - Ts - \sum_k \mu_k X_k \right),
\end{equation}
where $X_k$ is the global atomic fraction of element $k$, $\mu_k$ is the chemical potential of $k$, and $s$ is entropy, which in the large-system limit becomes
    \begin{equation}
    \label{entropy}
    s = k_{\mathrm{B}} \left( - \sum_k X_k \log_e X_k - \frac{1}{n} \sum_j P_j
    \log_e (P_j/\tilde{P}_{j})\right)
    \end{equation}
Minimizing the thermodynamic potential subject to the constraints $\sum_j P_j = 1$ and $\sum_j P_j x_{kj} = X_k$, where $x_{kj}$ is the atomic fraction of element $k$ in configuration $j$, gives the temperature-dependent population vector (see Figure \ref{fig:ltvc}(b))
\begin{equation} \label{population}
  P_j=\frac{\tilde{P}_{j}\,e^{-n\beta\left(\epsilon_j-\sum_k\mu_k x_{kj}\right)}}{{\sum_j \tilde{P}_{j}\,e^{-n\beta\left(\epsilon_j-\sum_k\mu_kx_{kj}\right)}}},
\end{equation}
where $\beta=1 / k_{\mathrm B} T$.

In the high-temperature limit, this becomes the temperature-independent population vector
\begin{equation} \label{po}
  \tilde{P_{j}}=\frac{g_j\,\prod_k X_k^{\,\,n \cdot \,x_{kj}}  }{{\sum\limits_{j' } g_{j'}\,\prod\limits_k X_k^{\,\,n\cdot\,x_{kj'}} }} ,
\end{equation}

{\bf iii.} Comparison of the behavior of the thermodynamic potentials and population vectors to experimental data leads to the introduction of two order parameters to estimate the transition temperature and miscibility gap.
The first order parameter is based on the angle between the population vector as a function of temperature
and the population vector for the fully disordered high temperature limit
\begin{equation}
\alpha\left({\bf X},T \right) \equiv {\bf P}\cdot {\bf \tilde{P}} / | {\bf P}|| {\bf \tilde{P}}|.
\end{equation}
The maximum of the rate of change of this angle as a function of temperature, max($\partial \alpha / \partial T$), is found to correspond to the order-disorder transition temperature at equi-atomic concentration (Figure \ref{fig:ltvc}(c)).
The second order parameter is the relative entropy (\emph{Kullback-Leibler} divergence \cite{RE1})
\begin{equation}
D_{\mathrm{KL}} \equiv \frac{1}{n} \sum_j P_j \log_e (P_j/\tilde{P}_{j})
\end{equation}
which is the entropy-loss due to ordering.
The surface of constant relative entropy is used to extrapolate the transition temperature at equi-atomic concentration to other compositions, and thus generate the order-disorder transition
boundary (Figure \ref{fig:ltvc}(d)) to delineate the miscibility gap in the phase diagram.

Defining a solid-solution-forming alloy as a composition where the order-disorder transition temperature is lower than the melting temperature (as obtained from experimental data or CALPHAD predictions),
58 out of 117 ($\sim 49.6\%$) investigated binary alloy systems are expected to form solid solutions --- 56 of these were corroborated by experimental reports.
148 out of 441 investigated ternary systems were predicted to form solutions ($\sim 33.6\%$), while 46\% of the 1100 investigated quaternaries and 130 quinaries are expected to form high-entropy alloys.

\section*{Vibrational Contributions}

\noindent
The works reviewed above focus on the role of configurational entropy in phase formation.
To understand the role of vibrations in phase formation in multi-component systems, the \AFLOWPOCC\ (\underline{P}artial \underline{Occ}upation) \cite{curtarolo:art110} and \AFLOWAPL\ (Automatic Phonon Library) \cite{curtarolo:art180} modules were combined to calculate the contribution of vibrations to the Gibbs free energy of formation for 3 recently synthesized \cite{curtarolo:art140, curtarolo:art148, curtarolo:art175, curtarolo:art179, curtarolo:art187} high-entropy carbides: (HfNbTaTiV)C, (HfNbTaTiW)C, and (HfNbTaTiZr)C.
The results \cite{curtarolo:art180} showed that for (HfNbTaTiV)C and (HfNbTaTiZr)C, where all of the binary precursors were in the same rocksalt structure as the multicomponent phase, the net contribution of the vibrational energy to mixing was negligible.
However for (HfNbTaTiW)C, the WC precursor is not rocksalt, and the vibrational contribution was significant.
Therefore, vibrational contributions are likely to be most significant when precursors or decomposition products have different structures than the high-entropy phase.

\section*{Conclusion}

\noindent
Thermodynamic analyses of the \AFLOW\ database leads to the following general conclusions and predictions about the distribution of different types of phases in alloy systems: (i) disorder is inevitable for multi-component systems \cite{curtarolo:art152}; (ii) competition between ordered structures plays an important role in glass formation \cite{curtarolo:art112, curtarolo:art154}; (iii) spectral descriptors based on the energy distribution of ordered structures can be used to predict the synthesizability of high-entropy phases \cite{curtarolo:art140}; (iv) high-entropy alloy transition temperatures can be predicted from the rate of change of the population vector as a function of temperature \cite{curtarolo:art139}; (v) vibrational contributions are important when the components have different structures than the high-entropy phase \cite{curtarolo:art180}.

\section*{Acknowledgments}
We thank Drs. S. Divilov, H. Eckert, C. Oses, D. Hicks, M. Esters, M. Mehl, S. Griesemer, R. Friedrich and X. Campilongo for insightful discussions.
CT acknowledges support from National Science Foundation (DMR-2219788).
SC acknowledges support from ONR (N000142112515).
This work was supported by high-performance computer time and resources from the DoD
   High Performance Computing Modernization Program (Frontier).

\newcommand{\Ozolins}{Ozoli{\c{n}}{\v{s}}}

\end{document}